# Continuity of the Lyapunov exponent for quasiperiodic operators with analytic potential.

J. Bourgain [*][†]    and    S. Jitomirskaya [‡]

*Dedicated to David Ruelle and Yasha Sinai*

## 1 Introduction.

In this paper we study continuity of the Lyapunov exponent associated with 1D quasiperiodic operators. Assume $v$ real analytic on $\mathbb{T}$. Let $v : \mathbb{T} \to \mathbb{R}$. Consider an $SL_2(\mathbb{R})$ valued function

$$A(x, E) = \begin{pmatrix} v(x) - E & -1 \\ 1 & 0 \end{pmatrix}, \quad x \in \mathbb{T}. \tag{1.1}$$

Set

$$M_N(E, x, \omega) = \prod_{j=N}^{1} A(S^j x), \ Sx = x + \omega,$$

$$L_N(E, \omega) = \frac{1}{N} \int \log \|M_N(E, x, \omega)\| dx.$$

The Lyapunov exponent is defined by $L(E, \omega) = \lim_{N \to \infty} L_N(E) = \inf_N L_N(E, \omega)$ and exists by subadditivity.

Our main result is the following Theorem:

---
[*]Institute for Advanced Study Princeton, NJ 08540

[†]Department of Mathematics, University of Illinois Urbana-Champaign, Urbana, IL 61801-2975

[‡]Department of Mathematics, University of California, Irvine, California 92697

**THEOREM 1** *Assume v real analytic on* $\mathbb{T}$. *Then*

- $L(E, \omega)$ *is continuous in* $E$.

- *if* $\omega_0$ *is irrational,* $L(E, \omega)$ *is jointly continuous in* $(E, \omega)$ *at every* $(E, \omega_0)$.

**Remark.** $L(E, \omega)$ may be discontinuous in $\omega$ at every rational $\omega$, see below.

Matrices $M_N$ appear in the study of 1D Schrödinger operators

$$(H_x \Psi)(n) = \Psi(n+1) + \Psi(n-1) + v(S^n x)\Psi(n), \qquad (1.2)$$

as $N-$step transfer-matrices, and $L(E)$ has been a subject of a considerable investigation in this context. Recently there were several results on regularity in $E$ for quasiperiodic operators (1.2) with $Sx = x + \omega, x \in T^d$. For typical $\omega$ (more precisely satisfying a strong Diophantine condition of the form

$$\|k\omega\| > C(|k|\log(1+|k|)^A)^{-1} \qquad (1.3)$$

Goldstein and Schlag [10] proved Hölder regularity of $L(E)$ for $d = 1$ and certain weaker regularity for $d > 1$ in the regime $L > 0$ (see also [5], Ch. VII). Precise estimates on Hölder regularity for the almost Mathieu operator at high coupling are contained in [4]. For $L = 0$ some regularity also holds [5], Ch. VIII. For a review of results on continuity of $L$ in $E$ for strictly ergodic shifts over finite alphabet (and new result of this type for S a primitive substitution) see Lenz [17]. As far as $\omega$-dependence, there was a number of results on continuity of the spectrum (e.g., [1, 7, 3, 16, 13]), but continuity of $L$ was not addressed directly. It is however important, since various quantities, including $L(E)$ can sometimes be effectively estimated or even directly computed for periodic operators obtained from the rational approximants of $\omega$. Let $\sigma(H)$ denote the spectrum of $H$.

**COROLLARY 2** *For the almost Mathieu operator* $H_{\lambda,\omega,x}$ *given by (1.2) with* $v(S^n x) = \lambda \cos 2\pi(x + n\omega)$, *we have* $L(E) = \max(0, \log \frac{|\lambda|}{2})$ *for all* $E \in \sigma(H)$, *all* $\lambda$ *and all irrational* $\omega$.

**Proof:** Krasovsky [15] showed that for $E \in \sigma(H_{\lambda, \frac{p}{q}})$, $L(E, \frac{p}{q})$ converges as $q \to \infty$ to (but is not equal to) $\max(0, \log \frac{\lambda}{2})$. The result then follows from Theorem 1 and continuity of spectra [1]. ∎

We will also list another immediate corollary of Theorem 1:

**COROLLARY 3** *Suppose $v$ analytic. Then*

$$\limsup_{N \to \infty} \frac{1}{N} \log \|M_N(E, x, \omega)\| \leq L(E) \tag{1.4}$$

*uniformly in $x$ and $E$ in a compact set.*

**Proof**: Furman [8] proved uniformity of (1.4) in $x$ for any continuous cocycle on a uniquely ergodic system. The result follows then from continuity of $L$ in $E$ and compactness. ∎

This uniformity is important for various questions arising in the nonperturbative analysis of operators (1.2). For example, Corollary 3 immediately implies that the almost Mathieu operator has strong dynamical localization (see [9]) for any $\lambda > 2$ and $\omega$ satisfying $\|k\omega\| > c(\omega)(|k|^{r(\omega)})^{-1}$, that is throughout the regime of [12]. Strong dynamical localization was obtained in [9] for $\omega$ satisfying a strong Diophantine condition (1.3). The restriction on $\omega$ was needed there only for the uniformity of an upper bound, such as given in Corollary 3, which is now established for all $\omega$.

We note that the continuity issue, even in $E$ alone, is nontrivial, as $L$ considered as a function on $C(\mathbb{T}, SL_2(\mathbb{R}))$, with $\omega$ fixed, is discontinuous at $A(\cdot, E)$ for a dense set of $E$ in the spectrum of corresponding $H$ provided $L(E)$ is positive and either $\omega$ is Liouville or $v$ even (follows from a theorem of Furman [8] and a combination of [2, 14], see also a discussion in [17] and a related result in [11]). Moreover, the restriction of $L$ to $C(\mathbb{T}, M)$ where $M$ is any locally closed submanifold of $SL_2(\mathbb{R})$ such that $A$ takes values in $M$, is also discontinuous at all such $A$ [8].

The rest of the paper is devoted to the proof of Theorem 1. Section 2 contains a large deviation theorem, which is applied in Section 3 together with avalanche principle to obtain estimates on convergence. Those estimates allow to approximate $L$ with $2L_{2N} - L_N$, uniformly in $(\omega, E)$ sufficiently close to $(\omega_0, E_0)$, where $N$ is also uniform in $(\omega, E)$, provided $\omega_0$ is irrational and $L(\omega_0, E_0)$ is positive. This is done in Section 4, and the proof of Theorem 1 is completed there.

Our proof builds on many of the same ideas and techniques as the proof of the regularity of $L(E)$ in [5], Ch. VII. A few arguments taken directly from [5] are given here without explanation, and the reader is therefore referred to [5] for more detail.

All constants $c, C$ in what follows will depend, unless otherwise noted, only on $v$ and $E$, being uniform for $E$ in a bounded range. Same notations will be sometimes used for different such constants.

S.J. is grateful to I. Krasovsky for useful discussions. The work of J.B. was supported in part by NSF Grant DMS-9801013 and of S.J. by NSF Grant DMS-0070755.

## 2 Large deviations.

**LEMMA 4** *Let*
$$|\omega - \frac{a}{q}| < \frac{1}{q^2} \qquad (a, q) = 1. \tag{2.1}$$

*Let $0 < \kappa < 1$. Then, for appropriate $c > 0$ and $C < \infty$, for $N > C\kappa^{-2}q$,*

$$\mathrm{mes}\,[x\,\big|\,\big|\frac{1}{N}\log\|M_N(E, x)\| - L_N(E)\big| > \kappa] < e^{-c\kappa q}. \tag{2.2}$$

**Proof.** Put
$$u(x) = \frac{1}{N}\log \|M_N(E; x)\| = \sum_{k \in \mathbb{Z}} \hat{u}(k) e^{2\pi i k x}$$

where
$$\hat{u}(0) = L_N(E)$$

As shown in [6] (see also [5], Ch. IV)

$$|\hat{u}(k)| < \frac{C}{|k|}. \tag{2.3}$$

Function $u$ also satisfies

$$|u(x)| < C \quad \text{and} \quad |u(x) - u(x + \omega)| < \frac{C}{N} \tag{2.4}$$

(see [5] for details.) Take
$$R \sim \kappa^{-1} q \qquad (2.5)$$
and, using (2.4), estimate for $N > C\kappa^{-2} q$
$$|u(x) - \sum_{|j|<R} \frac{R - |j|}{R^2} u(x + j\omega)| < C \frac{R}{N} < \frac{\kappa}{10}. \qquad (2.6)$$
for an appropriate $C$. Considering the Féjèr average, we obtain therefore
$$|u(x) - \hat{u}(0)| < \frac{\kappa}{10} + \sum_{0<|k|\leq K} |\hat{u}(k)| \, (1 + (R\|k\omega\|)^2)^{-1} + \alpha(x) \qquad (2.7)$$
where
$$\|\alpha\|_2^2 \leq \sum_{|k|>K} |\hat{u}(k)|^2 < C \sum_{k>K} \frac{1}{k^2} \sim K^{-1}. \qquad (2.8)$$
We estimate the second term in (2.7) as
$$\sum_{0<|k|<\frac{q}{4}} |k|^{-1} (2R\|k\omega\|)^{-1} + \sum_{\ell=1}^{4Kq^{-1}} \frac{1}{\ell q} \sum_{k \in I_\ell} (1 + (R\|k\omega\|)^2)^{-1} = (I) + (II)$$
where $I_\ell = [\ell \frac{q}{4}, (\ell+1)\frac{q}{4})$.

It follows from (2.1) that for $|k| \leq \frac{q}{2}, |k\omega - \frac{ka}{q}| < \frac{1}{2q}$ and hence $\|k\omega\| > \frac{1}{2q}$. Let $\alpha_1, \ldots, \alpha_{q/4}$ be the decreasing rearrangement of $(\|k\omega\|^{-1})_{0<k\leq \frac{q}{4}}$. Then we have $\alpha_i \leq \frac{2q}{i}$. Moreover, if $I$ is any interval of length $q/4$, same is true for $(\|k\omega\|^{-1}), k \in I$, if we exclude at most one value of $k$.

Hence, for an appropriate choice of $R$ in (2.5),
$$(I) \leq CR^{-1} \sum \frac{1}{k} \frac{q}{k} < CqR^{-1} < \frac{\kappa}{10},$$
and, for each $\ell$
$$\sum_{k \in I_\ell} (1 + (R\|k\omega\|)^2)^{-1} \leq 1 + \sum_{s=1}^{q} \left(R\frac{s}{q}\right)^{-2} \leq 1 + C\left(\frac{q}{R}\right)^2 \leq C$$
and
$$(II) < C \sum_{\ell=1}^{4Kq^{-1}} \frac{1}{\ell q} < Cq^{-1} \log K.$$
Letting $\log K \sim \kappa q, (II) < \frac{\kappa}{10}$, and (2.8) implies (2.2). ∎

# 3 Applications of Avalanche principle

Avalanche principle [10] (full details is also given in [5], Ch. VI) is the following:

Let $A_1, \cdots, A_n$ be a sequence in $SL_2(\mathbb{R})$ such that

$$\|A_j\| \geq \mu \tag{3.1}$$

$$\mu > n \tag{3.2}$$

and

$$|\log \|A_j\| + \log \|A_{j+1}\| - \log \|A_{j+1} A_j\|| < \frac{1}{2} \log \mu, \ j = 1, \ldots, n. \tag{3.3}$$

Then

$$\left| \log \left\| \prod_{j=n}^{1} A_j \right\| + \sum_{j=2}^{n-1} \log \|A_j\| - \sum_{j=1}^{n-1} \log \|A_{j+1} A_j\| \right| < C \frac{n}{\mu} \tag{3.4}$$

where $C$ is an absolute constant. We will also need the following extension, relaxing condition (3.2):

**LEMMA 5** *Assume* $A_1, \cdots, A_N$ *satisfy (3.1),(3.3) with* $\mu$ *sufficiently large and* $N = \prod_{i=1}^{s} n_i$ *where* $3 \leq n_i < \frac{\mu}{2}$, $i = 1, \ldots, s-1$ *and* $n_s < \mu$. *Then*

$$\left| \log \left\| \prod_{j=N}^{1} A_j \right\| + \sum_{j=2}^{N-1} \log \|A_j\| - \sum_{j=1}^{N-1} \log \|A_{j+1} A_j\| \right| < C_1 \frac{N}{\mu} \tag{3.5}$$

**Remarks:**

1. We show (3.5) with $C_1 = 5C$, $C$ from the avalanche principle. As will be seen from the proof, $C_1 = (3 + \epsilon)C$ will also work for large $\mu$.

2. The largeness condition on $\mu$ is explicit. For example, it is sufficient to have $\mu \log \mu > 27C$ with $C$ from (3.4).

**Proof:** For the sake of less cumbersome notations our proof will assume $N = 3^s$. The proof for the general case is exactly the same with obvious changes.

We use induction in $s$ with the beginning provided by (3.4) with $n = n_1, 2n_1$. Set $N_1 = 3^{s-1}$; $B_i = A_{3i} A_{3i-1} A_{3i-2}$, $i = 1, \ldots, N_1$. Then, by (3.4), for all $j$,

$$\left| \log \|B_j\| + \log \|A_{3j-1}\| - \log \|A_{3j-1} A_{3j-2}\| - \log \|A_{3j} A_{3j-1}\| \right| < \frac{3C}{\mu} \quad (3.6)$$

and similarly for $\log \|B_{i+1} B_i\|$.

(3.6) and (3.3) imply

$$\log \|B_j\| > \sum_{k=0}^{2} \log \|A_{3j-k}\| - \frac{3C}{\mu} - \log \mu > 2 \log \mu - \frac{3C}{\mu} = \log \mu_1 \quad (3.7)$$

where $\mu_1 > \mu$. Also,

$$|\log \|B_j\| + \log \|B_{j+1}\| - \log \|B_{j+1} B_j\|| <$$
$$\frac{12C}{\mu} + |\log \|A_{3j}\| + \log \|A_{3j+1}\| - \log \|A_{3j+1} A_{3j}\|| <$$
$$\frac{12C}{\mu} + \frac{1}{2} \log \mu < \frac{1}{2} \log \mu_1 \quad (3.8)$$

(it is for the last inequality in (3.8) that we need a largeness condition on $\mu$.)

Therefore, induction applies, and

$$\left| \log \left\| \prod_{j=N}^{1} A_j \right\| + \sum_{j=2}^{N_1-1} \log \|B_j\| - \sum_{j=1}^{N_1-1} \log \|B_{j+1} B_j\| \right| < C_1 \frac{3^{s-1}}{\mu_1} \quad (3.9)$$

Using (3.6) for each $B_j$ and $B_{j+1} B_j$ in (3.9) we obtain, after collecting terms,

$$\left| \log \left\| \prod_{j=N}^{1} A_j \right\| + \sum_{j=2}^{N-1} \log \|A_j\| - \sum_{j=1}^{N-1} \log \|A_{j+1} A_j\| \right| <$$
$$\frac{C_1 3^{s-1}}{\mu_1} + \frac{3C(3^{s-1} - 2)}{\mu} + \frac{6C(3^{s-1} - 1)}{\mu} < \frac{C_1 3^s}{\mu} \quad (3.10)$$

if $C_1 = 5C$. ∎

In case of positive Lyapunov exponent, large deviation theorem provides us a posibility to apply avalanche principle to $M_N(x + jN\omega)$ for $x$ in a set of large measure and therefore pass on to a larger scale.

**LEMMA 6** Let $\omega$ satisfy (2.1) and $L(E) > 100\kappa > 0$. Let $N > C\kappa^{-2}q$. Assume further $L_{2N}(E) > \frac{9}{10}L_N(E)$.

Then for $N_1$ s.t. $N|N_1$ and $N_1 N^{-1} = m < e^{c\kappa q}$, we have

$$|L_{N_1}(E) + \frac{m-2}{m}L_N(E) - 2\frac{m-1}{m}L_{2N}(E)| < C_1 e^{-c\kappa q}. \qquad (3.11)$$

**Remark.** Here $C$ is same as before, and $c$ is equal to $\frac{c}{2}$ from the large deviation theorem.

**Proof.** Apply the avalanche principle with

$$A_j = M_N(x + jN\omega, E)$$

and with $x$ restricted to the set $\Omega \subset \mathbb{T}$, s.t. for all $j \leq m$

$$\left|\tfrac{1}{N}\log\|M_N(E; x + jN\omega)\| - L_N(E)\right| < \kappa \qquad (3.12)$$

$$\left|\tfrac{1}{2N}\log\|M_{2N}(E; x + jN\omega)\| - L_{2N}(E)\right| < \kappa.$$

Thus from (2.2) and choice of $m$

$$\mathrm{mes}(\mathbb{T}\backslash\Omega) < 2me^{-c\kappa q} < Ce^{-\frac{c}{2}\kappa q}. \qquad (3.13)$$

Since $\|A_j\| > e^{N(L_N(E)-\kappa)} > e^{\frac{99}{100}NL_N}$ and $|\log\|A_j\| + \log\|A_{j+1}\| - \log\|A_{j+1}A_j\|| < 4N\kappa + 2N|L_N(E) - L_{2N}(E)| < \frac{6}{25}NL_N$, the avalanche principle applies. Thus, for sufficiently large $N$,

$$\left|\log\left\|\prod_{j=m}^{1} A_j\right\| + \sum_{j=2}^{m-1}\log\|A_j\| - \sum_{j=1}^{m-1}\log\|A_{j+1}A_j\|\right| < me^{-\frac{1}{2}NL_N}.$$

Integrating on $\Omega$, we get

$$\left|\int_\Omega \log\|M_{N_1}(E;x)\| + \sum_{j=2}^{m-1}\int_\Omega \log\|M_N(E; x+j\omega)\| - \sum_{j=1}^{m-1}\int_\Omega \log\|M_{2N}(E; x+j\omega)\|\right|$$
$$< me^{-\frac{1}{2}NL_N}.$$

Therefore, recalling (3.13)

$$|L_{N_1}(E) + \frac{m-2}{m}L_N(E) - \frac{2(m-1)}{m}L_{2N}(E)| < \frac{m}{N_1}e^{-\frac{1}{2}NL_N} + Ce^{-\frac{c}{2}\kappa q} < C_1 e^{-\frac{c}{2}\kappa q},$$

as claimed.

Lemma 6 may be iterated to get the following fact

**LEMMA 7** *Same assumptions as in Lemma 6.*
Then
$$|L_{N'} + L_N - 2L_{2N}| < e^{-c'\kappa q} + C\frac{N}{N'} \qquad (3.14)$$

holds for all $N'$ with $N|N'$ and $\frac{N'}{N} < \exp\exp\frac{c}{2}\kappa q$.

**Proof.** (3.14) follows from (3.11) with $c' = c$ if $N' < e^{c\kappa q}N$. Thus we may assume $N' > e^{c\kappa q}N$. Take $N_1 \sim e^{c\kappa q}N$ in order to apply Lemma 6. Thus

$$|L_{N_1} + L_N - 2L_{2N}| < Ce^{-c\kappa q} \qquad (3.15)$$

and

$$|L_{2N_1} + L_N - 2L_{2N}| < Ce^{-c\kappa q}$$

implying in particular

$$|L_{2N_1} - L_{N_1}| < 2Ce^{-c\kappa q}. \qquad (3.16)$$

Replacing $N$ by $N_1$ and taking $N_2 \sim e^{c\kappa q}N_1$, we get similarly from Lemma 6

$$|L_{N_2} + L_{N_1} - 2L_{2N_1}| < Ce^{c\kappa q} \qquad (3.17)$$

$$|L_{2N_2} - L_{N_2}| < 2Ce^{-c\kappa q}$$

and from (3.16), (3.17)

$$|L_{N_2} - L_{N_1}| < 5Ce^{-c\kappa q}. \qquad (3.18)$$

Letting in general $N_s \sim e^{c\kappa q}N_{s-1}$, we obtain

$$|L_{N_s} + L_{N_{s-1}} + L_{2N_{s-1}}| < Ce^{-c\kappa q} \qquad (3.19)$$
$$|L_{2N_s} - L_{N_s}| < 2Ce^{-c\kappa q} \qquad (3.20)$$
$$|L_{N_s} - L_{N_{s-1}}| < 5Ce^{-c\kappa q}. \qquad (3.21)$$

Consequently, from (3.18), (3.21)

$$|L_{N_s} - L_{N_1}| < 5Cse^{-c\kappa q}$$

and by (3.15)
$$|L_{N_s} + L_N - 2L_{2N}| < 6Cse^{-c\kappa q}. \qquad (3.22)$$

To get (3.14) with $c' = \frac{c}{2}$, we may allow $s < e^{\frac{c}{2}\kappa q}$ in (3.22), hence the estimate holds for $N'$ as stated. ∎

Lemma 7 will be sufficient for our induction step provided there exists an approximant $q$ with $e^{q_s} < q < \exp\exp c\kappa q_s$, where $q_s$ is the sequence of canonical rational approximants of $\omega$. For when this is not the case we need an additional statement.

**LEMMA 8** *In addition to the assumptions of Lemma 6, assume that $q|N$, $N < e^{c''\kappa q}$. Then*
$$|L_{N'} + L_N - 2L_{2N}| < C_2 e^{-\frac{c}{2}\kappa q} \tag{3.23}$$
*for all $5e^q N \leq N' \leq e^{-\frac{3}{2}q}q'$ with $N' = 3^s N$ or $\frac{N'}{2} = 3^s N$, where $q'$ is the next approximant after $q$.*

**Remark:** We assumed $N' = a3^s N$, $a = 1, 2$ for simplicity of formulation only. Lemma 8 holds as well for all $N'$ as in Lemma 5 with $\mu = e^{N(L_N - 2\kappa)}$.

**Proof:** The set $\Omega_N = \{x \in \mathbb{T} \mid |\frac{1}{N}\log\|M_N(E;x)\| - L_N(E)| > \kappa\}$ satisfies by (2.2) the measure estimate
$$\text{mes } \Omega_N < e^{-c\kappa q}. \tag{3.24}$$

Let us consider $v' = \sum_{|k|\leq N^2} \hat{v}(k)e^{2\pi ikx}$ a trigonometric polynomial of degree $N^2$, and let $M'_K$, $L'_K$, and $\Omega'_K$ be corresponding objects defined with $v$ replaced by $v'$.

Since
$$|\|M_N(x)\| - \|M'_N(x)\|| \leq \sup_x |v - v'|C^N,$$
we have that if $x \in \mathbb{T}\setminus(\Omega'_N \cup \Omega'_{2N})$, then
$$|\log\|M_K(x)\| - KL_K| < 2\kappa K, \ K = N, 2N, \tag{3.25}$$

However, $\Omega'_N$ admits a semi-algebraic description, therefore $\Omega'_N$ may be covered by at most $N^C$ intervals of size $< e^{-c\kappa q}$. The same holds for $\Omega'_N \cup \Omega'_{2N}$.

Hence, because of our upper bound on $N$, there is a collection $\mathcal{J}$ of at most $N^C$ intervals $I \subset \mathbb{T}$ s.t.
$$\text{mes }(\mathbb{T}\setminus \bigcup_{I \in \mathcal{J}} I) < e^{-\frac{c}{2}\kappa q} \tag{3.26}$$

and if $x \in \bigcup_{I \in \mathcal{J}} I, |x - x'| < e^{-q}$, then $x' \in \mathbb{T}\backslash(\Omega'_N \cup \Omega'_{2N})$, therefore $x'$ satisfies (3.25). Observe next that since $|\omega - \frac{a}{q}| < \frac{1}{qq'}$ and $q|N$,

$$\|\ell N\omega\| < \frac{\ell e^{c''\kappa q}}{qq'} < e^{-q}$$

for

$$\ell < e^{-\frac{3}{2}q}q'. \tag{3.27}$$

Hence, fixing $x \in \bigcup_{I \in \mathcal{J}} I$, it follows from the preceding that $x' = x + \ell N\omega$ will satisfy (3.25) for all $\ell$ as in (3.27).

Denoting

$$A_\ell = M_N(x + \ell N\omega)$$

we have thus

$$\|A_\ell\| > e^{N(L_N - 2\kappa)} > e^{\frac{49}{50}NL_N} \tag{3.28}$$

$$|\log \|A_\ell\| - NL_N| < 2\kappa N \tag{3.29}$$

$$|\log \|A_{\ell+1}A_\ell\| - 2NL_N| < (6\kappa + 2|L_{2N} - L_N|)N < \tfrac{13}{50}NL_N \tag{3.30}$$

since $L_N > 100\kappa$.

Therefore, for $N'$ as in the Lemma, we may now aplly Lemma 5 to obtain

$$\left| \frac{1}{N'} \log \Big\| \prod_{\ell = \frac{N'}{N}-1}^{0} A_\ell \Big\| + \frac{1}{N'} \sum_{\ell=2}^{\frac{N'}{N}-1} \log \|A_\ell\| - \frac{1}{N'} \sum_{\ell=1}^{\frac{N'}{N}-1} \log \|A_{\ell+1}A_\ell\| \right| < C_1 e^{-98\kappa N} < \frac{1}{5}e^{-q}. \tag{3.31}$$

Integrating (3.31) in $x \in \bigcup_{I \in \mathcal{J}} I$, and recalling (3.26) and the lower bound on $N'$, we get

$$|L_{N'} + L_N - 2L_{2N}| < e^{-q} + Ce^{-\frac{c}{2}\kappa q} < C_2 e^{-\frac{c}{2}\kappa q} \tag{3.32}$$

■

## 4  Proof of Theorem 1.

Assume $q_0$ is an approximant of $\omega$, thus

$$\left|\omega - \frac{a_0}{q_0}\right| < \frac{1}{q_0^2}, \qquad (a_0, q_0) = 1 \tag{4.1}$$

and $L(E) > 100\kappa > 0$. Here $\kappa$ is a small constant and we assume $q_0 > \kappa^{-2}$.

The construction below is described assuming $\omega \notin \mathbb{Q}$ but, as the reader will easily see, applies equally well for $\omega \in \mathbb{Q}$. In particular, the conclusion stated in Prop. 9, is valid in either case.

Since $100\kappa < L(E) \leq L_{2N} \leq L_N < C$ for any $N$, we may choose $N_0$, depending only on $\kappa, C, q_0$, satisfying

$$L_{2N_0} > \frac{9}{10} L_{N_0} \tag{4.2}$$

and

$$C\kappa^{-2} q_0 < N_0 < \kappa^{-C} q_0. \tag{4.3}$$

Set $q_{-1} = 0$. Starting from $q_0, N_0$, we construct a sequence of approximants $\{q_s\}$ of $\omega$ and integers $\{N_s\}$ such that

$$q_0 < N_0 < q_1 < \cdots < N_s < q_{s+1} < N_{s+1} < \cdots \tag{4.4}$$
$$q_{s+1} > e^{q_s} \tag{4.5}$$
$$C\kappa^{-2} q_s < N_s \sim q_s \text{ and, for } s \geq 1, N_{s-1} | N_s \tag{4.6}$$
$$|L_{N_{s+1}} + L_{N_s} - 2L_{2N_s}| < e^{-c_1 \kappa q_s} \tag{4.7}$$
$$|L_{2N_s} - L_{N_s}| < Ce^{-c_2 \kappa q_{s-1}} \tag{4.8}$$
$$|L_{N_{s+1}} - L_{N_s}| < e^{-c_3 \kappa q_{s-1}} \tag{4.9}$$

where $c' \gg c_1 > c_2 > c_3 > 0$. ($c' > 0$ the constant from Lemma 7).

Denoting $q_{s+1} > e^{q_s}$ the smallest approximant of $\omega$ satisfying (4.5), we distinguish 2 cases.

**Case I**: $q_{s+1} < e^{10 q_s}$

Take $N_{s+1}$ satisfying (4.6), hence $e^{q_s} < N_{s+1} < e^{11q_s} N_s$. Since $|\omega - \frac{a_s}{q_s}| < \frac{1}{q_s^2}$ and $N_s$ satisfies (4.6), (4.8), Lemma 7 applies with $q = q_s$, $N = N_s, N' = N_{s+1}$. Thus from (3.14)

$$|L_{N_{s+1}} + L_{N_s} - 2L_{2N_s}| < e^{-c' \kappa q_s} + e^{-\frac{1}{2} q_s} < 2e^{-c' \kappa q_s} < e^{-c_1 \kappa q_s} \tag{4.10}$$

and similarly with $N_{s+1}$ replaced by $2N_{s+1}$.

From (4.8), (4.10)

$$|L_{N_{s+1}} - L_{N_s}| < 2e^{-c' \kappa q_s} + 2Ce^{-c_2 \kappa q_{s-1}} < e^{-c_3 \kappa q_{s-1}}$$

and also
$$|L_{2N_{s+1}} - L_{N_{s+1}}| < 4e^{-c'\kappa q_s} < e^{-c_2\kappa q_s}.$$

**Case II**: $q_{s+1} \geq e^{10q_s}$.

Take again $N_{s+1}$ satisfying (4.6).

In this situation, we may not be able to apply Lemma 7 immediately and we perform some intermediate steps. Denote $q_s \leq q \leq e^{q_s}$ the approximant preceding $q_{s+1}$ and consider a first intermediate scale

$$N \sim \max(\kappa^{-2}q, e^{5c_1\kappa q_s}), q|N. \tag{4.11}$$

Thus, as in case (I)

$$|L_N + L_{N_s} - 2L_{2N_s}| < e^{-c'\kappa q_s} + e^{-4c_1\kappa q_s} < 2e^{-4c_1\kappa q_s} \tag{4.12}$$

and

$$|L_{2N} + L_{N_s} - 2L_{2N_s}| < 2e^{-4c_1\kappa q_s}. \tag{4.13}$$

A second scale $N'' \geq N$ is introduced as follows

If $q_{s+1} \leq e^{4q}$, let $N'' = N$.

If $q_{s+1} > e^{4q}$, let $N'' \sim e^{-2q}q_{s+1}$, with $N'' = 3^b N$. In the second case, we have conditions of Lemma 8 satisfied, and therefore (3.23) holds for both $N''$ and $2N''$. Therefore, we also have:

$$|L_{N''} - L_{2N''}| < Ce^{-\frac{c}{2}\kappa q}. \tag{4.14}$$

Next, apply Lemma 7 with $N = N''$ and $N' = N_{s+1} < C\kappa^{-2}e^{2q}N''$. Thus

$$|L_{N_{s+1}} + L_{N''} - 2L_{2N''}| < e^{-c'\kappa q} + C\frac{N''}{N_{s+1}} < 2e^{-c'\kappa q} \tag{4.15}$$

and similarly with $N_{s+1}$ replaced by $2N_{s+1}$.

Collecting the estimates (4.12), (4.13), (3.23) with $N' = N'', 2N''$, and (4.15), we obtain that

$$|L_{N_{s+1}} + L_{N_s} - 2L_{2N_s}| < 6e^{-4c_1\kappa q_s} + Ce^{-\frac{c}{2}\kappa q} + 2e^{-c\kappa q} < e^{-c_1\kappa q_s}$$

and similarly with $N_{s+1}$ replaced by $2N_{s+1}$. Therefore, in both cases I, II, (4.7) holds. (4.8) and (4.9) are then obtained as in case (I). This completes the construction.

As a consequence of (4.7) with $s = 0$ and (4.9)

$$|L(E) + L_{N_0}(E) - 2L_{2N_0}(E)| <$$
$$|L_{N_1} + L_{N_0} - 2L_{2N_0}| + \sum_{s \geq 1} |L_{N_{s+1}} - L_{N_s}| < e^{-c_1 \kappa q_0} + \sum_{s \geq 0} e^{-c_3 \kappa q_s} < 2e^{-c_3 \kappa q_0}.$$

Observe also that the assumption $L(E) > 100\kappa > 0$ in the beginning of this section could have been replaced by an assumption

$$L_N(E) > 100\kappa$$

for some $N$ chosen at least $\kappa^{-C} q_0$, $C$ some constant, as it is sufficient for the existence of $N_0$ satisfying (4.2),(4.3).

The conclusion is the following

**PROPOSITION 9** *Assume* $|\omega - \frac{a}{q}| < \frac{1}{q^2}$, $0 < \kappa < \frac{1}{100}$, $q > C\kappa^{-2}$ *and* $L_N(\omega, E) > \kappa$ *for some* $N > \kappa^{-C} q$. *Then there is* $N_0 < \kappa^{-C} q$, *and depending only on* $\kappa, C, q$, *s.t.*

$$|L(E) + L_{N_0}(E) - 2L_{2N_0}(E)| < e^{-c\kappa q}. \tag{4.16}$$

We can now finish the proof of the first statement of Theorem 1.

We may assume $\omega \notin \mathbb{Q}$. If $E_\alpha \to E$, then always, by subharmonicity, $\limsup L(E_\alpha) \leq L(E)$. We may therefore assume $L(E) > \kappa > 0$. Let $q > C\kappa^{-2}$ be an approximant of $\omega$. Taking $N > \kappa^{-C} q$, we have $L_N(E) > \kappa$ and hence also $L_N(E_\alpha) > \kappa$ for $\alpha > \alpha_0$. One may then choose $N_0$ s.t. (4.16) holds for both $E$ and $E_\alpha$. Thus

$$|L(E) - L(E_\alpha)| \leq |L_{N_0}(E) - L_{N_0}(E_\alpha)| + 2|L_{2N_0}(E) - L_{2N_0}(E_\alpha)| + 2e^{-c\kappa q}$$
$$\leq C(\kappa)^q |E - E_\alpha| + 2e^{-c\kappa q}$$

Thus $\limsup_\alpha |L(E) - L(E_\alpha)| \leq 2e^{-c\kappa q}$ and, letting $q \to \infty$, the result follows. To prove the second statement of Theorem 1, we assume $(\omega_\alpha, E_\alpha) \to (\omega_0, E_0)$. Note that since for each $N$, $L_N(\omega, E)$ is a subharmonic function in both variables, therefore, $L(\omega, E) = \inf_N L_N(\omega, E)$ is upper semicontinuous, so $\limsup_\alpha L(\omega_\alpha, E_\alpha) \leq L(\omega_0, E_0)$. Therefore we may assume $L(\omega_0, E_0) > \kappa > 0$. Let $q > C\kappa^{-2}$ be an approximant of $\omega$, hence

$$\left|\omega_0 - \frac{a}{q}\right| < \frac{1}{q^2}.$$

Taking again $N > \kappa^{-C}q$, we have $L_N(\omega_0, E_0) > \kappa$, hence $L_N(\omega_\alpha, E_\alpha) > \kappa$ and $|\omega_\alpha - \frac{a}{q}| < \frac{1}{q^2}$ for $\alpha > \alpha_0$.

Fixing any $\alpha > \alpha_0$, we may find $N_0 < \kappa^{-C}q$ s.t.

$$|L(\omega_0, E_0) + L_{N_0}(\omega_0, E_0) - 2L_{2N_0}(\omega_0, E_0)| < e^{-c\kappa q}$$

and

$$|L(\omega_\alpha, E_\alpha) + L_{N_0}(\omega_\alpha, E_\alpha) - 2L_{2N_0}(\omega_\alpha, E_\alpha)| < e^{-c\kappa q}.$$

Hence

$$|L(\omega_0, E_0) - L(\omega_\alpha, E_\alpha)| < C(\kappa)^q(|\omega_0 - \omega_\alpha| + |E_0 - E_\alpha|) + 2e^{-c\kappa q}$$
$$\limsup_\alpha |L(\omega_0, E_0) - L(\omega_\alpha, E_\alpha)| < 2e^{-c\kappa q}.$$

Letting $q \to \infty$, it follows that $L(\omega_0, E_0) = \lim_\alpha L(\omega_\alpha, E_\alpha)$. ∎